%% file: code-model-attack.tex
\documentclass[sigconf]{acmart}
\def \toolname {\textit{ALERT} }
\usepackage{comment}
\usepackage{listings}
\usepackage{xcolor}
\usepackage{balance}
\usepackage{tcolorbox}
\usepackage{listings}
\definecolor{mGreen}{rgb}{0,0.6,0}
\definecolor{mGray}{rgb}{0.5,0.5,0.5}
\definecolor{mPurple}{rgb}{0.58,0,0.82}
\definecolor{backgroundColour}{rgb}{0.95,0.95,0.92}

\lstdefinestyle{CStyle}{
    backgroundcolor=\color{backgroundColour},   
    commentstyle=\color{mGreen},
    keywordstyle=\color{magenta},
    numberstyle=\tiny\color{mGray},
    stringstyle=\color{mPurple},
    basicstyle=\footnotesize,
    breakatwhitespace=false,         
    breaklines=true,                 
    captionpos=b,                    
    keepspaces=true,                 
    numbers=left,                    
    numbersep=5pt,                  
    showspaces=false,                
    showstringspaces=false,
    showtabs=false,                  
    tabsize=2,
    language=C
}
\usepackage{enumitem}
\usepackage{multirow}
\usepackage[ruled,linesnumbered]{algorithm2e}
\usepackage{bm}
\usepackage{url}
\usepackage[colorinlistoftodos]{todonotes}
\usepackage{subfigure}
\SetAlFnt{\small}
\SetAlCapFnt{\small}
\SetAlCapNameFnt{\small}
\usepackage{algorithmic}
\algsetup{linenosize=\small}
\SetKwInput{KwInput}{Input}                % Set the Input
\SetKwInput{KwOutput}{Output}              % set the Output

\newboolean{showcomments}
\setboolean{showcomments}{true}
\ifthenelse{\boolean{showcomments}}
{ }
\ifthenelse{\boolean{showcomments}}
{ }
\ifthenelse{\boolean{showcomments}}
{ }
\AtBeginDocument{%
  \providecommand\BibTeX{{%
    \normalfont B\kern-0.5em{\scshape i\kern-0.25em b}\kern-0.8em\TeX}}}

\copyrightyear{2022} 
\acmYear{2022} 
\setcopyright{acmcopyright}\acmConference[ICSE '22]{44th International Conference on Software Engineering}{May 21--29, 2022}{Pittsburgh, PA, USA}
\acmBooktitle{44th International Conference on Software Engineering (ICSE '22), May 21--29, 2022, Pittsburgh, PA, USA}
\acmPrice{15.00}
\acmDOI{10.1145/3510003.3510146}
\acmISBN{978-1-4503-9221-1/22/05}
\begin{document}

\title{Natural Attack for Pre-trained Models of Code}

\author{Zhou Yang, Jieke Shi, Junda He and David Lo}

\affiliation{%
  \institution{School of Computing and Information Systems}
  \country{Singapore Management University}
}
\email{{zyang, jiekeshi, jundahe, davidlo}@smu.edu.sg}


\renewcommand{\shortauthors}{Zhou Yang, Jieke Shi, Junda He and David Lo.}

\begin{abstract}
  Pre-trained models of code have achieved success in many important software engineering tasks. However, these powerful models are vulnerable to adversarial attacks that slightly perturb model inputs to make a victim model produce wrong outputs. Current works mainly attack models of code with examples that preserve {\em operational program semantics} but ignore a fundamental requirement for adversarial example generation: perturbations should be natural to {\em human judges}, which we refer to as {\em naturalness} requirement.

  In this paper, we propose \toolname (N\textbf{a}tura\textbf{l}n\textbf{e}ss Awa\textbf{r}e A\textbf{t}tack), a black-box attack that adversarially transforms inputs to make victim models produce wrong outputs. Different from prior works, this paper considers the {\em natural} semantic of generated examples at the same time as preserving the {\em operational} semantic of original inputs. 
  Our user study demonstrates that human developers consistently consider that adversarial examples generated by \toolname are more natural than those generated by the state-of-the-art work by Zhang et al. that ignores the naturalness requirement.
  On attacking CodeBERT, our approach can achieve attack success rates of 53.62\%, 27.79\%, and 35.78\% across three downstream tasks: vulnerability prediction, clone detection and code authorship attribution. On GraphCodeBERT, our approach can achieve average success rates of 76.95\%, 7.96\% and 61.47\% on the three tasks. The above outperforms the baseline by 14.07\% and 18.56\% on the two pre-trained models on average. Finally, we investigated the value of the generated adversarial examples to harden victim models through an adversarial fine-tuning procedure and demonstrated the accuracy of CodeBERT and GraphCodeBERT against {\em ALERT}-generated adversarial examples increased by 87.59\% and 92.32\%, respectively.
\end{abstract}

\begin{CCSXML}
<ccs2012>
  <concept>
       <concept_id>10011007.10011074.10011099.10011102.10011103</concept_id>
       <concept_desc>Software and its engineering~Software testing and debugging</concept_desc>
       <concept_significance>300</concept_significance>
  </concept>
  <concept>
  <concept_id>10010147.10010257.10010293.10010294</concept_id>
  <concept_desc>Computing methodologies~Neural networks</concept_desc>
  <concept_significance>500</concept_significance>
  </concept>
<concept>
  <concept_id>10011007.10011074.10011784</concept_id>
  <concept_desc>Software and its engineering~Search-based software engineering</concept_desc>
  <concept_significance>500</concept_significance>
  </concept>
</ccs2012>
\end{CCSXML}

\ccsdesc[500]{Software and its engineering~Software testing and debugging}  
\ccsdesc[500]{Computing methodologies~Neural networks}
\ccsdesc[500]{Software and its engineering~Search-based software engineering}

\keywords{Genetic Algorithm, Adversarial Attack, Pre-Trained Models}


\maketitle

\input{sections/intro.tex}
\input{sections/preliminary.tex}

\input{sections/methodology.tex}
\input{sections/experiment.tex}

\input{sections/result.tex}

\input{sections/discussion.tex}
\input{sections/related_work.tex}
\input{sections/conclusion.tex}

\begin{acks}
This research was supported by the Singapore Ministry of Education (MOE) Academic Research Fund (AcRF) Tier 1 grant.
\end{acks}

\balance
\bibliographystyle{ACM-Reference-Format}
\bibliography{reference}

\end{document}

%% file: sections/intro.tex
\section{Introduction}

% Recently, researchers have created pre-trained models of code~\cite{CodeBERT, GraphCodeBERT} that can boost the performance on programming language processing tasks. Feng et al. \cite{CodeBERT} train CodeBERT on a corpus containing both natural and programming languages, and results show that CodeBERT can achieve outstanding performance on code search and code documentation generation. GraphCodeBERT \cite{GraphCodeBERT}, which additionally considers structural information of source code (i.e., data flow), outperforms CodeBERT on four downstream tasks. 

Recently, researchers~\cite{while_box_attack_code,9438605,MHM} have shown that models of code like {\em code2vec}~\cite{code2vec} and {\em code2seq}~\cite{code2seq}, can output different results for the two code snippets sharing the same operational semantics, one of which is generated by renaming some variables in the other. The modified code snippets are called {\em adversarial examples}, and the models under attack are called {\em victim models}.

\begin{figure*}[t!]
    \centering
    \subfigure[An original code snippet that can be correctly classified by a model fine-tuned on CodeBERT.]{
    \includegraphics[width=0.266\linewidth]{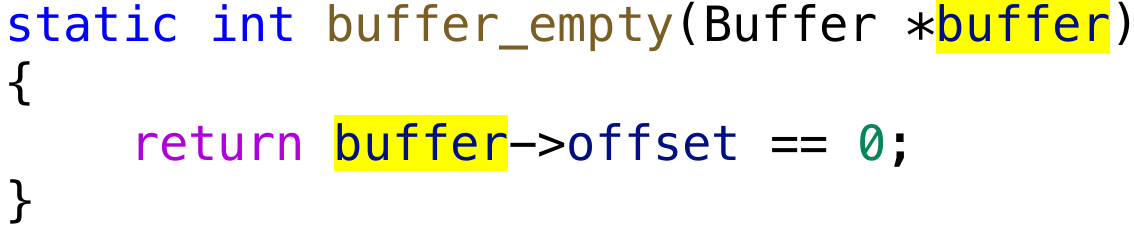}
    \label{fig:original}
    }
    \quad
    \subfigure[MHM generates an adversarial example by replacing the variable {\tt buffer} to {\tt qmp\_async\_cmd\_handler}.]{
        \includegraphics[width=0.383\linewidth]{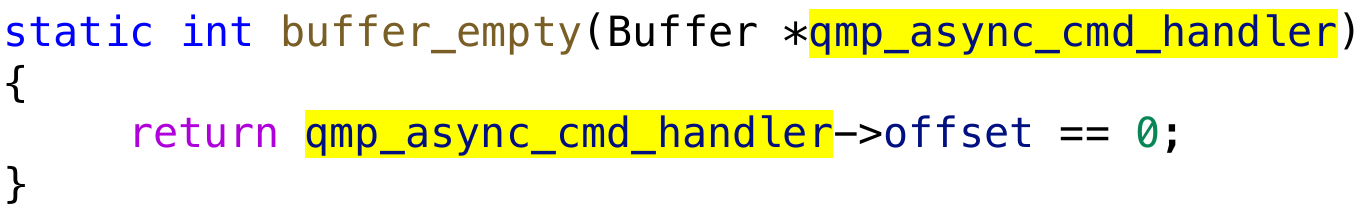}
    \label{fig:mhm}
    }
    \quad
    \subfigure[\toolname generates an adversarial example by replacing the variable {\tt buffer} to {\tt queue}.]{
        \includegraphics[width=0.2626\linewidth]{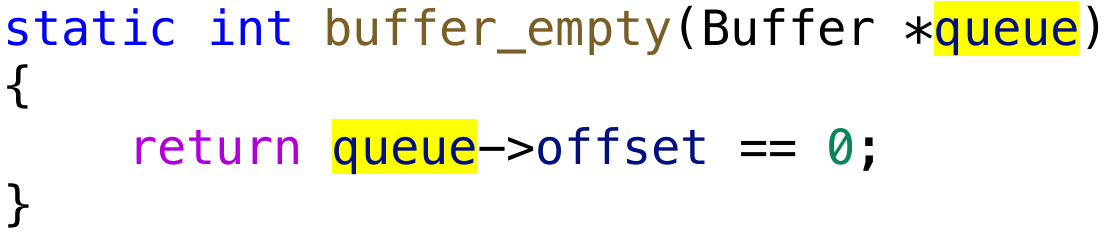}
    \label{fig:ga}
    }
    \caption{The original example in (a) is from the dataset used in Zhou et al.’s study~\cite{zhou_devign_2019}. Both MHM and {\em ALERT} can generate successful adversarial examples by substituting a variable name. But HMH uses an unnatural replacement while \toolname uses a more natural replacement that can better fit into the context and more closely relates to the original variable name.}
	\label{fig:example}
\end{figure*}

Naturalness is a fundamental requirement in adversarial example generation. For example, perturbations to images are constrained with the infinity norm to ensure naturalness~\cite{FGSM, PGD}. Attack for NLP models also requires adversarial examples to be fluent and natural~\cite{Jin_Jin_Zhou_Szolovits_2020}. We propose that the naturalness requirement is also essential for attacking models of code. 
Casalnuove et al.~\cite{dual_channel} provide a dual-channel view of source code: machines that compile and execute code mainly focus on the operational semantics, while developers often care about natural semantics of code (e.g., names of variables) that can assist human comprehension.
Although many automated tools have been included into the software development process, there is no doubt that software development is still a process led by humans. Code that violates coding convention or has poor variable names, may be acceptable for machines but rejected by humans. For example, Tao et al. \cite{6976093} report that 21.7\% of patches in Eclipse and Mozilla projects were rejected because the patches used bad identifier names or violated coding conventions.
As a result, unnatural adversarial examples may not even pass code reviews and not to mention being merged into codebases. 

Existing works on attacking models of code are effective~\cite{9438605,MHM,while_box_attack_code}, but they focus on preserving operational semantics and barely pay attention to whether adversarial examples are natural to human judges. 
For instance, the state-of-the-art black-box method, MHM~\cite{MHM}, randomly selects replacements from a fixed set of variable names without considering semantic relationships between original variables and their substitutes. 
Figure \ref{fig:mhm} shows an adversarial example generated by replacing the variable name {\verb|buffer|} in Figure \ref{fig:original} to {\verb|qmp_async_cmd_handler|}. Even though the new program preserves the operational semantics, {\verb|qmp_async_cmd_handler|} is not a natural replacement of {\tt buffer} to human judges considering the context (surrounding code). The natural semantics (i.e., human understanding) of {\tt buffer} clearly do not overlap with {\verb|qmp_async_c|}\\{\verb|md_handler|}. In this paper, we argue that adversarial examples for models of code should consider preserving the semantics at two levels: operational semantics (catering for machines as audience) and natural semantics (catering for humans as audience).

The neglect of naturalness requirements in current attack methods motivates us to propose \toolname (N\textbf{a}tura\textbf{l}n\textbf{e}ss Awa\textbf{r}e A\textbf{t}tack), a black-box attack that is aware of natural semantics when generating adversarial examples of code. Similar to MHM~\cite{MHM}, \toolname renames variables to generate adversarial examples.
%Some other works, e.g., \cite{9438605, wang2019coset, Ramakrishnan2020}, consider other types of transformation (e.g., converting {\tt switch} statements to {\tt if} statements); however, they have not been shown to achieve as high attack success rates as MHM. Moreover, they are usually rule-based: they are less risky to affect the naturalness of code as they do not modify variable names with specific meanings. 
Our approach has three main parts: a natural perturbation generator, a heuristic method that tries to generate adversarial examples as fast as possible, and a genetic algorithm-based method to search for adversarial examples more comprehensively in case the heuristic method fails.

This paper investigates the victim models that are fine-tuned on the state-of-the-art pre-trained models, CodeBERT~\cite{CodeBERT} and GraphCodeBERT~\cite{GraphCodeBERT}.
\toolname first uses the {\em masked language prediction} function in pre-trained models to generate natural substitutes. Given a code snippet with several masked tokens, this function can utilize the context information to predict the potential values of masked tokens. We leverage such a function in CodeBERT and GraphCodeBERT to generate candidate substitutes for each variable. Then, to pick the substitutes that are semantically closer, we use pre-trained models to compute contextualized embeddings of these new tokens and calculate its Cosine similarity (for measuring the semantic distances)~\cite{10.5555/1394399} with embeddings of the original tokens. We rank these candidates according to Cosine similarities and only select the top-$k$ candidates as natural substitution candidates.

\toolname has two steps to search adversarial examples using natural substitution candidates. It first uses a greedy algorithm (Greedy-Attack) and then applies a genetic algorithm (GA-Attack) if the former fails. The Greedy-Attack defines a metric to measure the {\em importance} of variable names in a code snippet and starts to substitute variables with the highest importance. An algorithm guided by the importance can find successful adversarial examples faster than the random sample strategy used in MHM~\cite{MHM}. When substituting a variable, Greedy-Attack greedily selects the replacement (out of all {\em natural substitutes}), from which the generated adversarial example makes the victim model produce lower confidence on the ground truth label. If it fails to change the prediction results, Greedy-Attack continues to replace the next variable until all the variables are considered or an adversarial example is obtained. But generating adversarial examples for code is essentially a combinatorial problem, and the greedy algorithm may generate sub-optimal results. If the greedy algorithm fails, we use the GA-Attack to perform a more comprehensive search.

We first conduct a user study to examine whether searching from substitutes generated by \toolname can produce adversarial examples that are natural to human judges. Participants give a {\em naturalness score} (1 for very unnatural and 5 for very natural) to each adversarial example. Results show that participants consistently provide a higher score to {\em ALERT}-generated examples (on average 3.95) than examples generated by the original MHM~\cite{MHM} (on average 2.18) that selects substitutes randomly over all variables.

Then, we evaluate MHM and \toolname on the six victim models (2 pre-trained models $\times$ 3 tasks). We consider three relevant tasks that may be adversely affected by such an attack: vulnerability prediction, clone detection and code authorship attribution.\footnote{For example, malicious users may write vulnerable code snippets and do not want them to be identified.} Since we argue that adversarial examples should look natural to human developers, we make MHM search on the same set of natural substitutes generated by {\em ALERT}.
On CodeBERT, \toolname can achieve attack success rates of 53.62\%, 27.79\%, and 35.78\% across three downstream tasks. MHM only reaches 35.66\%, 20.05\% and 19.27\%, respectively, which means that \toolname can improve attack success rates over MHM by 17.96\%, 7.74\% and 16.51\%. On GraphCodeBERT, our approach achieves success rates of 76.95\%, 7.96\% and 61.47\% on the same three tasks, outperforming MHM by 21.78\%, 4.54\%, and 29.36\%. Finally, we investigate the value of generating adversarial examples by using them to harden the models through an adversarial fine-tuning strategy. We demonstrate that the robustness of CodeBERT and GraphCodeBERT increased by 87.59\% and 92.32\% after adversarial fine-tuning with examples generated by {\em ALERT}. The contributions of this paper include:
\begin{itemize}[leftmargin=*]
    \item We are the first to highlight the naturalness requirement in generating adversarial examples for models of code. We also propose {\em ALERT} that is aware of natural semantics when generating adversarial variable substitutes. A user study confirms that using these substitutes can generate adversarial examples that look natural to human judges. \toolname can also achieve higher attack success rates than a previous method.
    \item We are the first to develop adversarial attacks on CodeBERT and GraphCodeBERT, and show that models fine-tuned on state-of-the-art pre-trained models are vulnerable to such attacks. 
    \item We show the value of {\em ALERT}-generated examples: adversarially fine-tuning victim models with these adversarial examples can improve the robustness of CodeBERT and GraphCodeBERT against \toolname by 87.59\% and 92.32\%, respectively.
\end{itemize}

The rest of this paper is organized as follows. Section~\ref{sec:preliminary} briefly describes preliminary materials. In Section~\ref{sec:Methodology}, we elaborate on the design of the proposed approach {\em ALERT}. We describe the settings of the experiment in Section~\ref{sec:experiment}, and present the results of our experiments that compare the performance of \toolname and some baselines in Section~\ref{sec:result}. After summarising the threats to validity in Section~\ref{sec:ttv}, Section~\ref{sec:related_work} discusses some related works. Finally, we conclude the paper and present future work in Section \ref{sec:conclusion}.

%% file: sections/preliminary.tex
\section{Preliminaries}
\label{sec:preliminary}
This section briefly introduces some preliminary information of this study, including pre-trained models of code, adversarial example generation for DNN models, and the Metropolis-Hastings Modifier (MHM) method that we use as the baseline.

\subsection{Pre-trained Models of Code}
\label{subsec:pre_train_models}
Pre-trained models of natural language like BERT \cite{bert} have brought breakthrough changes to many natural language processing (NLP) tasks, e.g., sentiment analysis~\cite{9240704}. 
Recently, researchers have created pre-trained models of code~\cite{CodeBERT, GraphCodeBERT} that can boost the performance on programming language processing tasks.

Feng et al. propose CodeBERT~\cite{CodeBERT} that shares the same model architecture as RoBERTa~\cite{RoBERTa}. CodeBERT is trained on a bimodal dataset (CodeSearchNet~\cite{CodeSearchNet}), a corpus consisting of natural language queries and programming language outputs. CodeBERT has two training objectives. One objective is \textit{masked language modeling} (MLM), which aims to predict the original tokens that are masked out in an input. The other objective is \textit{replaced token detection} (RTD), in which the model needs to detect which tokens in a given input are replaced. Experiment results have shown that in downstream tasks like code classification or code search, which requires understanding the code, CodeBERT could yield superior performance, although it is less effective in code-generation tasks.
GraphCodeBERT~\cite{GraphCodeBERT} also uses the same architecture as CodeBERT, but the former additionally considers the inherent structure of code, i.e., data flow graph (DFG). GraphCodeBERT keeps the MLM training objective and discards the RTD objective. It designs two DFG-related tasks: data ﬂow edge prediction and node alignment. GraphCodeBERT outperforms CodeBERT on four downstream tasks.

There are some other pre-trained models of code. CuBERT~\cite{CuBERT} is trained on Python source code and C-BERT~\cite{CBERT} is a model trained on the top-100 starred GitHub C language repositories. CodeGPT~\cite{CodeXGLUE} is a Transformer-based language model pre-trained on programming languages for code generation tasks. In this paper, we focus on analyzing CodeBERT and GraphCodeBERT, as they can work on multiple programming languages. Besides, recent studies~\cite{CodeXGLUE,zhouxin,chengransaner} have empirically shown that CodeBERT and GraphCodeBERT demonstrate state-of-the-art performance across multiple code processing tasks.

\subsection{Adversarial Example Generation}
Although Deep Neural Network (DNN) models have achieved great success on many tasks, many research works~\cite{sensei,yang2022revisiting} have shown that state-of-the-art models are vulnerable to adversarial attacks. Adversarial attacks aim to fool DNN models by slightly perturbing the original inputs to generate adversarial examples that are natural to human judges. Many techniques have been proposed to show that adversarial examples can be found for models in different domain, including, image classification~\cite{FGSM,PGD}, reinforcement learning~\cite{gleave2019adversarial,guo2021adversarial}, sentiment analysis~\cite{biasfinder}, speech recognition~\cite{audioattack}, machine translation~\cite{acl_NLP_attack}, etc. 

According to the information of victim models that an attacker can access, adversarial attacks can be divided into two types: \textit{white-box} and \textit{black-box}. In white-box settings, attackers can access all the information of the victim models, e.g., using model parameters to compute gradients. But white-box attacks often lack practicality since the victim models are usually deployed remotely (e.g., on cloud services), and typically attackers can only access the APIs to query models as well as corresponding outputs.
Black-box attacks mean that an attacker only knows the inputs and outputs of victim models (e.g., predicted labels and corresponding confidence). This paper proposes a novel black-box attack to mislead models that have the state-of-the-art performance. 

Adversarial attacks can also be categorized into \textit{non-targeted} attack and \textit{targeted} attack. Non-targeted attacks only aim to make a victim model produces wrong predictions, while targeted attacks force a victim model to make specific predictions. For example, a targeted attack may require a classifier to predict all the deer images as a horse while a non-target requires a classifier to predict an image incorrectly. The attack proposed in this paper is non-targeted.

\subsection{Metropolis-Hastings Modifier (MHM)}
\label{subsec:mhm}

Considering the fact that models can be remotely deployed so that model parameters are inaccessible, we focus on black-box attacks for models of code. This section introduces the baseline used in this paper.
Zhang et al.~\cite{MHM} formalizes the process of adversarial example generation as a sampling problem. The problem can be decomposed into an iterative process consisting of three stages: (1) selecting the variable to be renamed (2) selecting the substitutions and (3) deciding whether to accept to replace the variable with selected substitution. 

Zhang et al.~proposed Metropolis-Hastings Modifier (MHM)~\cite{MHM}, a Metropolis-Hastings sampling-based~\cite{Metropolis1953} identiﬁer renaming technique to solve this problem and generate adversarial examples for models of code.
This method is a black-box attack that randomly selects replacements for local variables and then strategically determines to accept or reject replacements. It uses both predicted labels and corresponding confidence of the victim model to select adversarial examples more effectively. 
MHM pre-defines a large collection of variable names, from which the replacements are selected. However, neither the creation of this collection nor selecting replacements considers the natural semantics. As a result, MHM produces examples that are not natural to human judgments. For example, suppose we change a variable name to an extremely long string that is not semantically close to the original variable. In that case, it may change the result of CodeBERT and GraphCodeBERT since the long name will be tokenized into multiple sub-tokens, impacting the output significantly. However, developers certainly will not accept this code.

In this paper, similar to MHM, we use variable renaming as the adversarial example generation technique and explore how to produce adversarial examples that are natural. We choose MHM~\cite{MHM} as our baseline as it does not require gradient information and also uses fine-grained model outputs (i.e., predicted results and corresponding confidence) to perform renaming and achieves good attack success rates (of one degree of magnitude higher than other black-box approaches~\cite{9438605}). For example, Pour et al.'s approach \cite{9438605} only causes an absolute decrease of 2.05\% to {\em code2vec}'s performance on the method name prediction task. To ensure that renamed variables make no changes in operational semantics, similar to MHM, we only rename local variables in code snippets.

%% file: sections/methodology.tex
\section{Methodology}
\label{sec:Methodology}
This paper proposes \toolname (N\textbf{a}tura\textbf{l}n\textbf{e}ss Awa\textbf{r}e A\textbf{t}tack), a black-box attack that leverages the pre-trained models that victim models are fine-tuned on.
It generates substitutes that are aware of natural semantics, which are called naturalness-aware substitutions in this paper. \toolname takes two steps to search for adversarial examples that are likely to be natural to human judges. The first step (Greedy-Attack) is optimized to find adversarial examples fast, and the second step (GA-Attack) is applied to do a more comprehensive search if the former fails.

\subsection{Naturalness-Aware Substitution}
\label{subsec:generation}
\toolname leverages the two functions of pre-trained models to generate and select naturalness-aware substitutes for variables: masked language prediction and contextualized embedding. To generate natural substitutes for one single variable (e.g., {\verb|index2dict|}), it operates in three steps:

\vspace{0.2cm}\noindent \textbf{Step 1}. We convert code snippets into a format that CodeBERT or GraphCodeBERT can take as inputs. Source code often contains many domain-specific abbreviations, jargon and their combinations, which are usually not included in the vocabulary set and cause the out-of-vocabulary problem~\cite{shi2022identifier,bigvocab}. Both CodeBERT and GraphCodeBERT use Byte-Pair-Encoding (BPE) \cite{Gage1994ANA,sennrich2016neural} to deal with such out-of-vocabulary problems by tokenizing a word into a list of sub-tokens. For example, a variable {\verb|index2dict|} can be converted into three sub-words ({\verb|index|}, {\verb|2|}, {\verb|dict|}) and then fed into the model.

\vspace{0.2cm}\noindent \textbf{Step 2}. Then, we generate potential substitutes for each sub-token. For the sake of simplicity but without any loss of generality, let us imagine a case where there is only one variable (e.g., {\verb|index2dict|}) that only appears once in an input. We use $T = \langle t_1, t_2, \cdots, t_m\rangle$ to represent the sequence of sub-tokens that BPE produces from the variable name. %For any sub-token in the sequence, the potential substitutes are all the words in the vocabulary set. 
For each sub-token in the sequence, we use the masked language prediction function of CodeBERT or GraphCodeBERT to produce a ranked list of potential substitute sub-tokens. %can predict masked tokens in an input. 
%They assign a probability to each word and uses the word with the highest value as the prediction for a position. 
Instead of just picking a single output, we select the top-$j$ substitutes. Intuitively, these substitutes are what pre-trained models think can fit the context better (compared to other sub-tokens). Still, not all of them are semantically similar to the original sub-tokens.

\vspace{0.2cm}\noindent \textbf{Step 3}. We assume that $\langle t_i, t_{i+1}$, $t_{i+2} \rangle$ is a sequence of sub-tokens of one variable name (e.g., corresponding to {\verb|index|}, {\verb|2|} and {\verb|dict|}). We replace the sub-tokens in the original sequence $T$ with candidate sub-tokens (e.g., $t_i', t_{i+1}', t_{i+2}'$) generated in Step 2 to get $T'$. After that, the pre-trained model computes the contextualized embeddings of each sub-token in $T'$, and we fetch the embeddings for $t_i', t_{i+1}'$ and $t_{i+2}'$. We concatenate these new embeddings and compute its Cosine similarity with concatenated embeddings of $t_i, t_{i+1}$ and $t_{i+2}$ in $T$. The cosine similarity is used as a metric to measure to what extent a sequence of candidate sub-tokens is similar to the original variable's sequence of sub-tokens. 
We rank the substitutes in descending order by the value of Cosine similarity.
In the end, we select top-$k$ sequences of substitute sub-tokens with higher similarity values and revert them into concrete variable names. 
%The process repeats for each substitutes.

One code snippet often contains multiple variables that appear in various positions. Algorithm \ref{algo:generation} displays how we apply the above process to each variable extracted from the source code. First, we use a parser to extract variable names ({\tt vars}) from the input ({\tt extract()} at Line 2) and then enumerate all the variables and their occurrences in the code (Line 3-4). The process discussed above is then applied to each variable occurrence to generate potential substitutes ({\tt perturb()} at Line 6). We take the union of the substitutes sets for all occurrences of a variable (Line 7). We then remove duplicated and invalid words, e.g., those that do not comply with the variable naming rules or those that are keywords in programming languages ({\tt filter()} at Line 9), after which we return filtered substitutes (Line 11). We refer to these filtered substitutes as the {\em naturalness-aware} substitutes.

\begin{algorithm}[t]
  \caption{Naturalness Aware Substitutes Generation} 
  \label{algo:generation}
  \SetAlgoLined
  \KwInput{$c$: input source code, $M$: pre-trained model}
  \KwOutput{$subs$: substitutes for variables}
  $subs$ = $\varnothing$\;
  $vars$ = extract($c$)\;
  \For(){
    $var$ {\em in} $vars$
    }{
        \For(){
            $occ$ {\em in} $var.occurrences$
        }{
            \# $var.occurrences$ returns all occurrences of $var$ \;
            $tmp\_subs$ = perturb($occ$, $c$,  $M$)\;
            $subs[var]$ = $subs[var] \bigcup tmp\_subs$\;
        }
        $subs[var]$ = filter($subs[var]$)\;
}

\algorithmicreturn{ $subs$}
\end{algorithm}

\subsection{Greedy-Attack}

\subsubsection{Overall Importance Score}
\label{subsubsec:OIS}
To perform semantic-preserving transformation by renaming variables, an attacker first needs to decide which tokens in a code snippet should be changed. Inspired by adversarial replacements for NLP tasks~\cite{bert-attack} that prioritizes more important tokens in a sentence, for each variable in a code snippet, we first measure its contribution to helping the model make a correct prediction. We introduce a metric called the importance score to quantify such contribution. Formally speaking, the importance score of the $i^{th}$ token in a code snippet $c$ is defined as follow:
\begin{equation}
  IS_i = M(c)[y] - M(c^*_{-i})[y]
\end{equation}

In the above formula, $y$ is the ground truth label for $c$ and $M(c)[y]$ represents the confidence of $M$'s output corresponding to the label $y$. A new code snippet generated by substituting variable names is called a {\em variant}. A variant $c^*_{-i}$ and is created by replacing the $i^{th}$ token (which must be a variable name) in $c$ with $\langle unk \rangle$, which means that the literal value at this position is unknown. Intuitively, the importance score approximates how knowing the value of the $i^{th}$ token affects the model's prediction on $c$. If $IS_i > 0$, it means that the token $t_i$ can help model make correct prediction on $c$. As stated in Section~\ref{subsec:generation}, one code snippet often contains multiple variables that appear in multiple positions. All the occurrences of a variable should be updated accordingly when performing adversarial attacks, so we extend the definition of importance score for a single token to the overall importance score (OIS) for a variable. OIS is computed as follow:
\begin{equation}
  OIS_{var} = \sum_{i \in var[pos]} IS_i
\end{equation}
where $var$ is a variable in $c$, and $var[pos]$ means all occurrences of $var$ in $c$. It is noticed that the definition of OIS can better reflect the unique property of attacking models of programming languages as compared to models of natural languages. Even though a variable at one position is trivial, appearing more often can make it an important variable (i.e., a vulnerable word) in adversarial attacks. The overall importance score can be viewed as an analogy to the gradient information in white-box attacks. For example, if the gradients are larger at some positions of inputs (e.g., certain pixels), then it is easier to change the model outputs if we perturb those positions.

Based on tree-sitter\footnote{\url{https://tree-sitter.github.io/tree-sitter/}}, a multi-language parser generator tool, we implement a name extractor that can retrieve all the variable names from syntactically valid code snippets written in C, Python or Java. More specifically, to avoid altering the operational semantics, we only extract the local variables that are defined and initialized within the scope of the code snippet and swap them with valid variable names that have never occurred in the code. To improve accuracy, variable names that collide with a field name are also excluded. After extraction, we compute the OIS for each variable and proceed to the next step.

\subsubsection{Word Replacement}
We design an OIS-based greedy algorithm to search substitutes that can generate adversarial examples. Algorithm~\ref{algo:greedy-attack} illustrates the process of this Greedy-Attack. First, we rank extracted variables from the original code snippet in descending order according to their OIS (Line 2 to 3). We select the first variable from them and find all its candidate substitutes generated following the process described in Section~\ref{subsec:generation} (Line 4 to 6). We replace the variable in the original input with these substitutes to create a list of variants, after which these variants are sent to query the victim model. We collect returned results and see if at least one variant makes the victim model make wrong predictions (Line 9 to 12). If there is such a variant, the Greedy-Attack returns it as a successful adversarial example. Otherwise, we replace the original input with the variant that can mostly reduce the victim model's confidence on the results and select the next variable to repeat the above processes (Line 15). Greedy-Attack terminates either when a successful adversarial example is found (Line 11) or when all the extracted variables are enumerated (Line 17). 

Considering OIS information is beneficial to the Greedy-Attack in two aspects. First, as discussed in Section~\ref{subsubsec:OIS}, if a variable has a higher OIS, it indicates significant impacts of modifying this variable in the code snippet. Giving higher priorities to variables with larger OIS can help find successful adversarial examples faster, which means that fewer queries to the victim model are required. It increases the usability of our attack in practice since remotely deployed black-box models often constrain the query frequency. Secondly, finding successful adversarial examples early also means fewer variables are modified in an original code snippet, making the generated adversarial examples more natural to human judges.

% \begin{figure}[t!]
% 	\centering
% 	\includegraphics[width=1\linewidth]{figures/fig2.pdf}
% 	\caption{Process of Greedy-Attack.}
% 	\label{fig:greedy_attack}
% \end{figure}

\begin{algorithm}[!t]
  \caption{Greedy-Attack Workflow} 
  \label{algo:greedy-attack}
  \SetAlgoLined
  \KwInput{$c$: input source code, $subs$: substitutes for variables in $c$}
  \KwOutput{$c'$: adversarial example}
  $c'$ = $c$\;
  $vars$ = $extract(c)$ \# extract $vars$ from $c$\;
  $vars$ = sort($vars$) \# sort $vars$ according to OIS\;
  \For(){
    $var$ {\em in} $vars$
    }{
        $list\_c$ = $\varnothing$\;
        \For(){
            $sub$ {\em in} $subs[var]$
        }{
            \# iterate all the substitutes for $var$\;
            $tmp\_c$ = replcae($c'$, $var$, $sub$)\;
        \If{$M(tmp\_c) \neq M(c)$}{
          $c' = tmp\_c$\;
          return $c'$\;
        }
            $list\_c$ = $list\_c \bigcup tmp\_c$\;
        }
        $c'$ = select($list\_c$) \# select the adversarial example with lowest model's confidence on the ground truth label\;}
  \algorithmicreturn{ $c'$}
\end{algorithm}

\subsection{GA-Attack}
Finding appropriate substitutes to generate adversarial examples is essentially a combinatorial optimization problem, whose objective is to find the optimal combination of variables and corresponding substitutes that minimizes the victim model's confidence on the ground truth label. Greedy-Attack can run faster but may be stuck in a single local optimal, leading to low attack success rates. We also design an attack based on genetic algorithms (GA), called GA-Attack. If the Greedy-Attack fails to find a successful adversarial example, we apply GA-Attack to search more comprehensively. Algorithm~\ref{algo:ga-attack} shows the overview of how GA-Attack works. It first initializes the population (Line 1, more detailed are given in Section~\ref{subsubsec:initailzation}), and then performs genetic operators to generate new solutions (Line 2 to 11). GA-Attack computes the fitness function (Section \ref{subsubsec:fitness}) and keep solutions with larger fitness values (Line 13). In the end, the algorithm returns the solution with the highest fitness value (Line 15 to 16).

\subsubsection{Chromosome Representation}
\label{subsubsec:representation}
In GA, the chromosome represents the solution to a target problem, and a chromosome consists of a set of genes. In this paper, each gene is a pair of an original variable and its substitution. GA-Attack represents chromosomes as a list of such pairs. For example, assuming that only two variables ({\verb|a|} and {\verb|b|}) can be replaced in an input program, the chromosome $\langle {\verb|a|}:{\verb|x|}, {\verb|b|}:{\verb|y|} \rangle$ means replacing {\verb|a|} to {\verb|x|} and {\verb|b|} to {\verb|y|}.

\subsubsection{Population Initialization}
\label{subsubsec:initailzation}
In the running of GA, a population (a set of chromosomes) evolves to solve the target problem. GA-Attack maintains a population whose size is the number of extracted variables that can be substituted. Since GA-Attack will be triggered only after Greedy-Attack fails, it can leverage the information discovered in the previous step. For each extracted variable, Greedy-Attack finds its substitution that can decrease the victim model's confidence on the ground truth label most. Given one variable and the substitution found by Greedy-Attack, GA-Attack creates a chromosome that only changes this variable to the substitution and keeps other variables unchanged. The process is repeated for each variable in a code snippet to obtain a population. For example, assuming that three variables ({\verb|a|}, {\verb|b|} and {\verb|c|}) are extracted from an input program, and Greedy-Attack suggests $\langle {\verb|a|}:{\verb|x|}, {\verb|b|}:{\verb|y|}, {\verb|c|}:{\verb|z|} \rangle$, GA-Attack initializes a population of three chromosomes: $\langle {\verb|a|}:{\verb|x|}, {\verb|b|}:{\verb|b|}, {\verb|c|}:{\verb|c|} \rangle$, $\langle {\verb|a|}:{\verb|a|}, {\verb|b|}:{\verb|y|}, {\verb|c|}:{\verb|c|} \rangle$. and $\langle {\verb|a|}:{\verb|a|}, {\verb|b|}:{\verb|b|}, {\verb|c|}:{\verb|z|} \rangle$.

\subsubsection{Operators}
Greedy-Attack runs in multiple iterations. In each iteration, two genetic operators (mutation and crossover) are used to produce new chromosomes (i.e., children). We apply crossover with a probability of $r$ and mutation with a probability of $1-r$ (Line 8). The mutation operation (Line 9) on two chromosome ($c_1$ and $c_2$) works as follows: we first randomly select a cut-off position $h$, and replace $c_1$'s genes after the position $h$ with $c_2$'s genes at the corresponding positions. As an example, for two chromosomes ($c_1 = \langle {\verb|a|}:{\verb|x|}, {\verb|b|}:{\verb|y|}, {\verb|c|}:{\verb|c|} \rangle$ and $c_2 = \langle {\verb|a|}:{\verb|x|}, {\verb|b|}:{\verb|b|}, {\verb|c|}:{\verb|z|} \rangle$) and a cut-off position $h = 2$, the child generated by crossover is $\langle {\verb|a|}:{\verb|x|}, {\verb|b|}:{\verb|y|}, {\verb|c|}:{\verb|z|} \rangle$. Given a chromosome in the population, the mutation operator randomly selects a gene and then replaces it with a randomly selected substitute. For instance, {\verb|a|} in $\langle {\verb|a|}:{\verb|x|}, {\verb|b|}:{\verb|b|} \rangle$ is selected and ${\verb|a|}:{\verb|x|}$ becomes ${\verb|a|}:{\verb|aa|}$.

\begin{algorithm}[!t]
  \caption{GA-Attack Workflow
  } 
  \label{algo:ga-attack}
  \SetAlgoLined
  \KwInput{$c$: input source code, $max\_iter$: max iteration, $r$: crossover rate, $child\_size$: number of generated children in each iteration}
  \KwOutput{$c'$: adversarial example}
  $population$ = greedy\_initialization($c$)\;
  \While{not exceed $max\_iter$}{
    {}
    $child\_list = []$ \;
    \While{$len(child\_lis) < child\_size$}{
        $p = \cup (0,1)$ \;
        \eIf{$p < r$}{
          $child$ = crossover($population$) \;
        }{
          $child$ = mutation($population$)\;
        }
        $child\_list$.append($child$)\;
    }
    $population$ = selection($population \cup child\_list$)\;
  }
  $c'$ = argmax(population); {\# select the one with highest fitness value}\\
  \algorithmicreturn{ $c'$}
\end{algorithm}

\subsubsection{Fitness Function}
\label{subsubsec:fitness}
GA uses a fitness function to measure and compare the quality of chromosomes in a population. A higher fitness value indicates that the chromosome (variable substitutions) is closer to the target of this problem. We compute the victim model's confidence values with respect to the ground truth label on the original input and the variant. The difference between confidence values is used as the fitness value. Assuming $T$ is the original input and $T'$ is a variant corresponding to a chromosome, the fitness value of this chromosome is computed by:
\begin{equation}
  fitness = M(T)[y] - M(T')[y]
\end{equation}

After generating children in one iteration, we merge them to the current population and perform a selection operator (Line 14). GA-Attack always maintains a population of the same size (i.e., numbers of extract variables). It discards the chromosomes that have lower fitness values.

%% file: sections/experiment.tex
\section{Experiment Setup}
\label{sec:experiment}

\subsection{Datasets and Tasks}
\label{subsec:dataset}
We introduce the three downstream tasks and their corresponding datasets used in our experiments.
The statistics of datasets are presented in Table~\ref{tab:model_performance}.

\subsubsection{Vulnerability Prediction}
This task aims to predict whether a given code snippet contains vulnerabilities. We use the dataset that was prepared by Zhou et al. \cite{zhou_devign_2019}. The dataset is extracted from two popular open-sourced C projects: FFmpeg\footnote{https://www.ffmpeg.org/} and Qemu\footnote{https://sites.google.com/view/devign}.
In Zhou et al.'s dataset, 27,318 functions are labeled as either containing vulnerabilities or clean. This dataset is included as part of the CodeXGLUE benchmark \cite{CodeXGLUE} that has been used to investigate the effectiveness of CodeBERT for vulnerability prediction. CodeXGLUE divides the dataset into training, development and test set that we reuse in this study.

\subsubsection{Clone Detection}
The clone detection task aims to check whether two given code snippets are clones, i.e., equivalent in operational semantics. 
BigCloneBench \cite{BigCodeBench} is a broadly recognized benchmark for clone detection, containing more than six million actual clone pairs and 260,000 false clone pairs from various Java projects. Each data point is a Java method. In total, the dataset has covered ten frequently-used functionalities. Following the settings of prior works~\cite{wang2020detecting,wei_supervised_2017}, we filtered the data which do not have a label and then balanced the dataset to make the ratio of true and false pairs to 1:1. To keep the experiment at a computationally friendly scale, we randomly select 90,102 examples for training and 4,000 for validation and testing.

\subsubsection{Authorship Attribution}
The authorship attribution task is to identify the author of a given code snippet. We did our experiments with the Google Code Jam (GCJ) dataset, which is originated from \textit{Google Code Jam} challenge\footnotetext{https://codingcompetitions.withgoogle.com/codejam}, a global coding competition that Google annually hosts. Alsulami et al.~\cite{alsulami_source_2017} collected the GCJ dataset and made it publicly available. The GCJ dataset contains 700 Python files (70 authors and ten files for each author), but we notice that some Python files are C++ code. After discarding these C++ source code files, we get 660 Python files in total. 20\% of files are used for testing, and 80\% of files are for training.

\subsection{Target Models}

\begin{table}[!t]
    \caption{Statistics of Datasets and of Victim Models.}
    \begin{tabular}{llll}
    \hline
    Tasks                                                                             & Train/Dev/Test       & Model         & Acc \\ \hline
    \multirow{2}{*}{\begin{tabular}[c]{@{}l@{}}Vulnerability\\ Prediction \cite{zhou_devign_2019}\end{tabular}} & \multirow{2}{*}{21,854/2,732/2,732}   & CodeBERT & 63.76\%     \\  &   & GraphCodeBERT & 63.65\%   \\ \hline
    \multirow{2}{*}{\begin{tabular}[c]{@{}l@{}}Clone\\ Detection \cite{BigCodeBench}\end{tabular}} & \multirow{2}{*}{90,102/4,000/4,000}  & CodeBERT      & 96.97\%     \\                     &       & GraphCodeBERT & 97.36\%   \\ \hline
    \multirow{2}{*}{\begin{tabular}[c]{@{}l@{}}Authorship\\ Attribution \cite{alsulami_source_2017}\end{tabular}} & \multirow{2}{*}{528/--/132}  & CodeBERT      & 90.35\%     \\     &          & GraphCodeBERT & 89.48\%   \\ \hline
\end{tabular}
    \label{tab:model_performance}
\end{table}

This paper investigates the robustness of the state-of-the-art pre-trained models, CodeBERT \cite{CodeBERT} and GraphCodeBert \cite{GraphCodeBERT}. To obtain the victim models, we fine-tune CodeBERT and GraphCodeBERT on the three tasks mentioned in Section \ref{subsec:dataset}.

\subsubsection{CodeBERT} CodeBERT~\cite{CodeBERT} is a pre-trained model that is capable of learning from bimodal data in the form of both programming languages and natural languages. When fine-tuning CodeBERT on vulnerability prediction and clone detection task, we use the same parameter settings adopted in the CodeXGLUE \cite{CodeXGLUE} except that we increase the maximal input length to 512 and achieve a slightly higher performance than results reported in the CodeXGLUE paper. Since there is no instruction on the hyper-parameter setting for fine-tuning on authorship attribution task, we use the same settings, and the obtained model can achieve 90.35\% accuracy, slightly higher than the accuracy of the LSTM model reported in \cite{alsulami_source_2017}.

\subsubsection{GraphCodeBERT} GraphCodeBERT~\cite{GraphCodeBERT} considers the inherent structure of the program and takes advantage of the data-flow representation. We set the maximal input length of GraphCodeBERT to 512 and follow the same setting for other hyper-parameters in the GraphCodeBERT paper~\cite{GraphCodeBERT} to fine-tune it on the three downstream tasks. On the clone detection task, the model can achieve an accuracy of 97.36\%, almost the same with the performance of 97.3\% reported in \cite{GraphCodeBERT}. On the vulnerability prediction and authorship attribution task, GraphCodeBERT also achieves the performance that is comparable with the results of CodeBERT.

The performance of these models is displayed in Table~\ref{tab:model_performance}. The results we obtain are closed to results reported in their original papers and another recent paper~\cite{CodeBERT, GraphCodeBERT, CodeXGLUE}, highlighting that the victim models used in our experiments are adequately fine-tuned.

\subsection{Settings of Attacks}
\toolname has a number of hyper-parameters to be set, including the number of natural substitutions generated for each variable and parameters for GA-Attack in Algorithm \ref{algo:ga-attack}. Our experiment setting allows \toolname to generate 60 candidate substitutions for each variable occurrence, and it selects the top 30 substitutions ranked by the cosine similarity with original embedding. For GA-Attack, we set $child\_size$ as 64 and set a dynamic value for the maximal iterations ($max\_iter$): the larger one of 5 times the number of extracted variables or 10. The crossover rate $r$ is set as $0.7$.

We consider MHM~\cite{MHM} as our baseline, which has two hyper-parameters: the maximum number of iterations and the number of variables sampled in each iteration. The MHM paper~\cite{MHM} suggests setting the latter as 30 but does not provide a standard setting for the maximal iterations. In each iteration, MHM needs to query the victim model many times, which is time-consuming. We sampled 5\% testing data from the vulnerability prediction task and found that over 95\% successful adversarial examples are found before 100 iterations. To make the MHM experiment within a computational friendly scale, we set the maximum number of iterations of MHM to 100. The original MHM can only perturb C programs, so we extend it to perturb Python and Java code.

%% file: sections/result.tex
\begin{table*}[!t]
    \caption{Comparison results of Attack Success Rates (ASR) on attacking CodeBERT and GraphCodeBERT across three tasks. The numbers in the parentheses correspond to the absolute improvement with respect to the attack success rates of MHM-NS.}
    \begin{tabular}{lllllllll}
        \hline
        \multirow{2}{*}{Task}   & \multicolumn{3}{l}{CodeBERT}                      & \multicolumn{3}{l}{GraphCodeBERT}                 \\ \cline{2-7} 
        & MHM-NS     & Greedy-Attack             & \toolname          & MHM-NS     & Greedy-Attack             & \toolname         \\ \hline
Vulnerability Detection & 35.66\% & 49.42\% (+13.76\%) & \bf{53.62\% (+17.96\%)} & 55.17\%   & 71.98\% (+16.81\%)  &\bf{76.95\% (+21.78\%)}     \\
Clone Detection         & 20.05\% & 23.20\% (+3.15\%)            & \bf{27.79\%  (+7.74\%)}         & 3.42\%    & 6.75\% (+3.33\%)               & \bf{7.96\% (+4.54\%)}              \\
Authorship Attribution  & 19.27\% & 30.28\% (+11.01\%)            & \bf{35.78\%  (+16.51\%)}        & 32.11\%    & 46.79\%  (+14.68\%)             & \bf{61.47\%  (+29.36\%)}             \\ \hline
Average  & 24.99\% & 34.30\% (+9.31\%)            & \bf{39.06\%  (+14.07\%)}        & 30.23\%    & 42.17\%  (+11.94\%)             & \bf{48.79\%  (+18.56\%)}  \\ \hline
        \end{tabular}
        \label{tab:ASR}
    \end{table*}
    
\section{Experiment Results and Analysis}
\label{sec:result}

In this section, we perform experiments to answer research questions related to the performance of adversarial attacks. We care about naturalness, attack success rates and scalability as well as the value of using adversarial examples to improve model robustness via adversarial fine-tuning, which are discussed by answering three research questions, respectively.

\subsection*{RQ1. How natural are the adversarial examples generated by \textit{ALERT}?}
\begin{figure}[!t]
	\centering
	\includegraphics[width=0.8\linewidth]{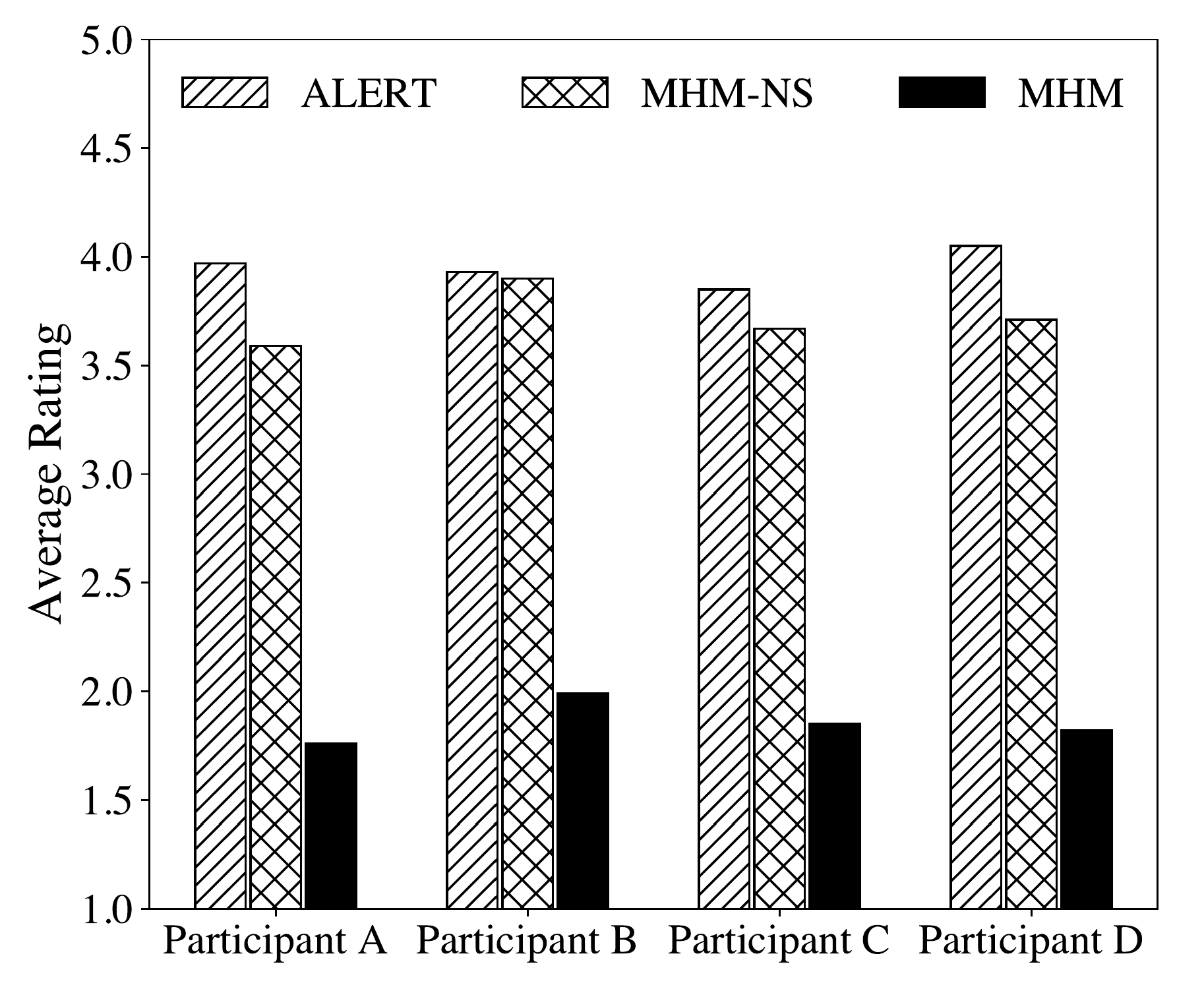}
	\caption{Results of the user study to evaluate naturalness of adversarial examples. The y-axis corresponds to the average ratings (5 means very natural; 1 means very unnatural). The x-axis represents distinguished independent participants.}
	\label{fig:user_study}
\end{figure}
When generating substitutions for variables in code, \toolname takes the natural semantics of adversarial examples into consideration.
This research question explores whether these naturalness-aware substitutions can help produce adversarial examples that are more natural to human judges. To answer this question, we conduct a user study to analyze the naturalness of examples generated by MHM, MHM-NS and the proposed \toolname method. Unlike the original MHM that ignores the naturalness, MHM-NS selects a replacement from the same pool of naturalness-aware substitutions as {\em ALERT}.

\begin{figure*}[t!]
    \centering
    \subfigure[VCR on attacking CodeBERT]{
    \includegraphics[width=0.22\linewidth]{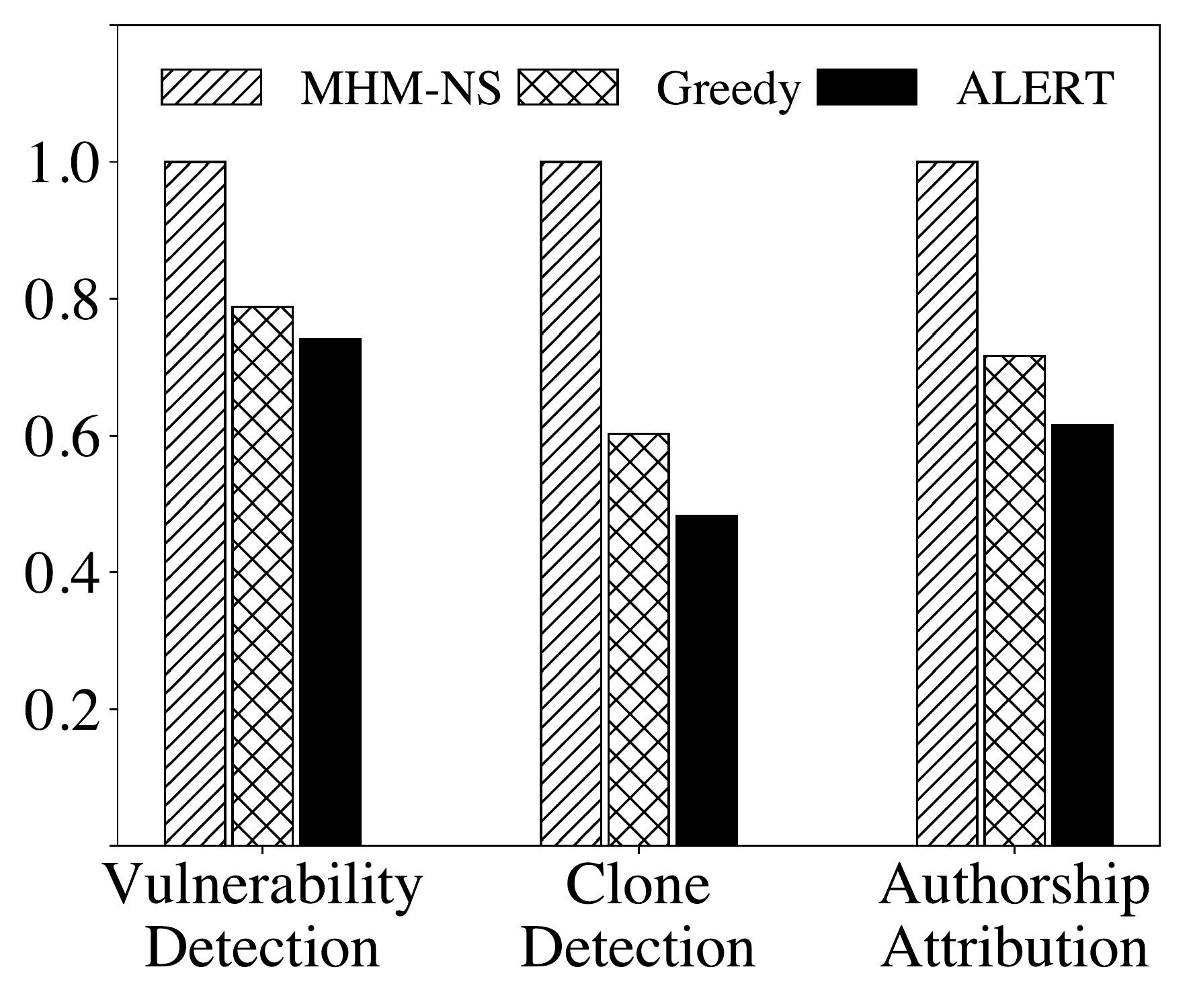}
    }
    \quad
    \subfigure[NoQ on attacking CodeBERT]{
        \includegraphics[width=0.22\linewidth]{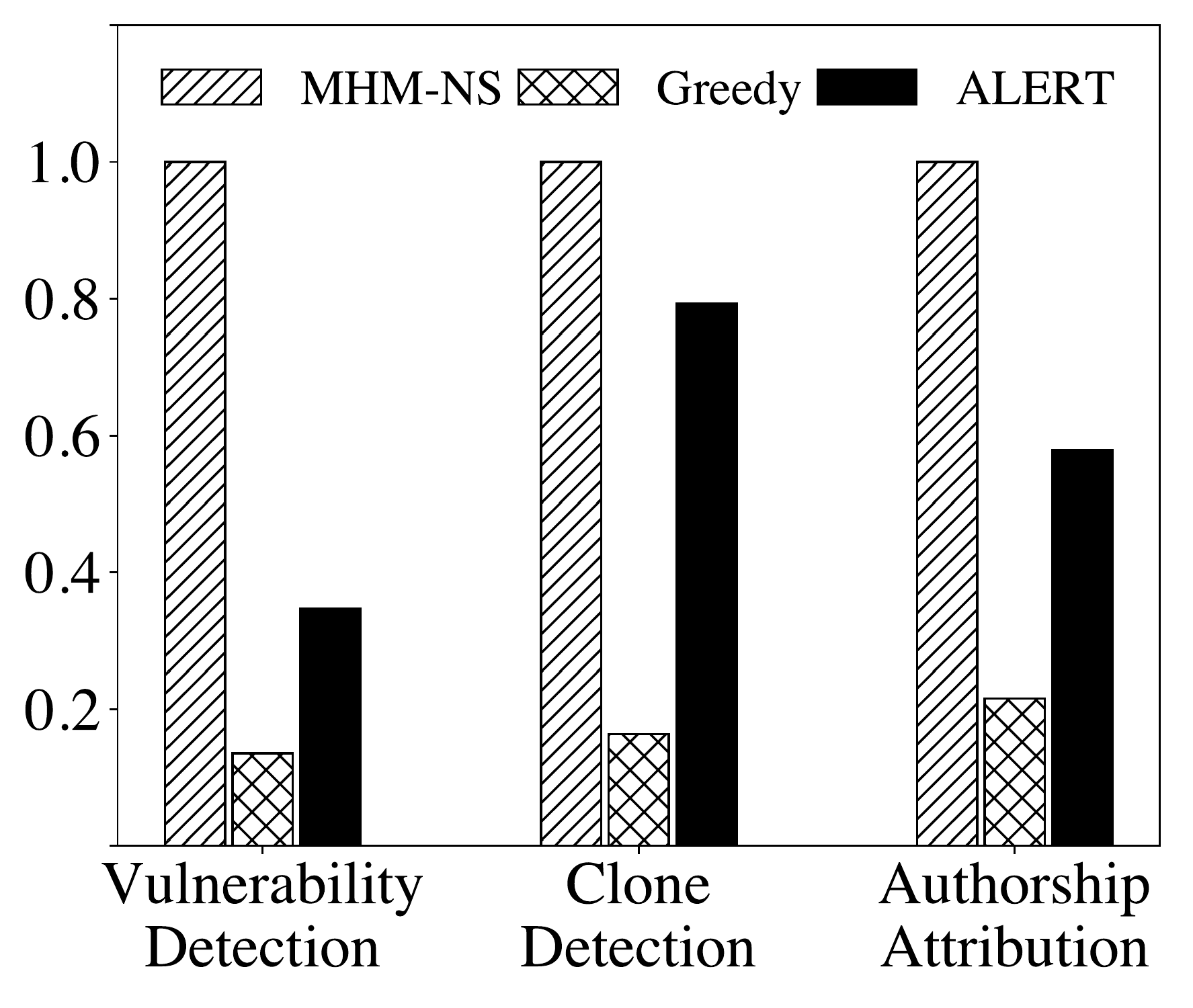}
    }
    \quad
    \subfigure[VCR on attacking GraphCodeBERT]{
        \includegraphics[width=0.22\linewidth]{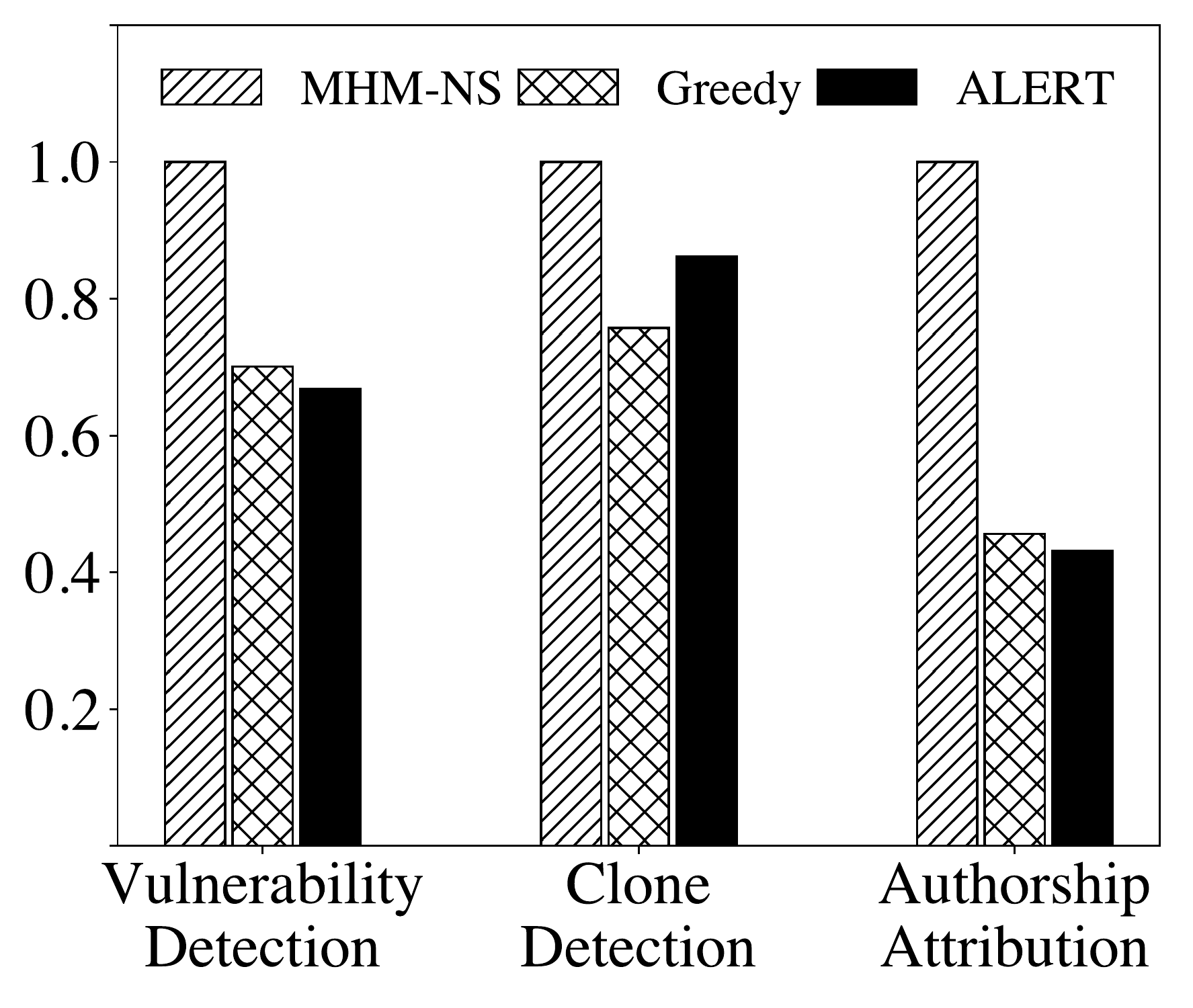}
    }
    \quad
    \subfigure[NoQ on attacking GraphCodeBERT]{
        \includegraphics[width=0.22\linewidth]{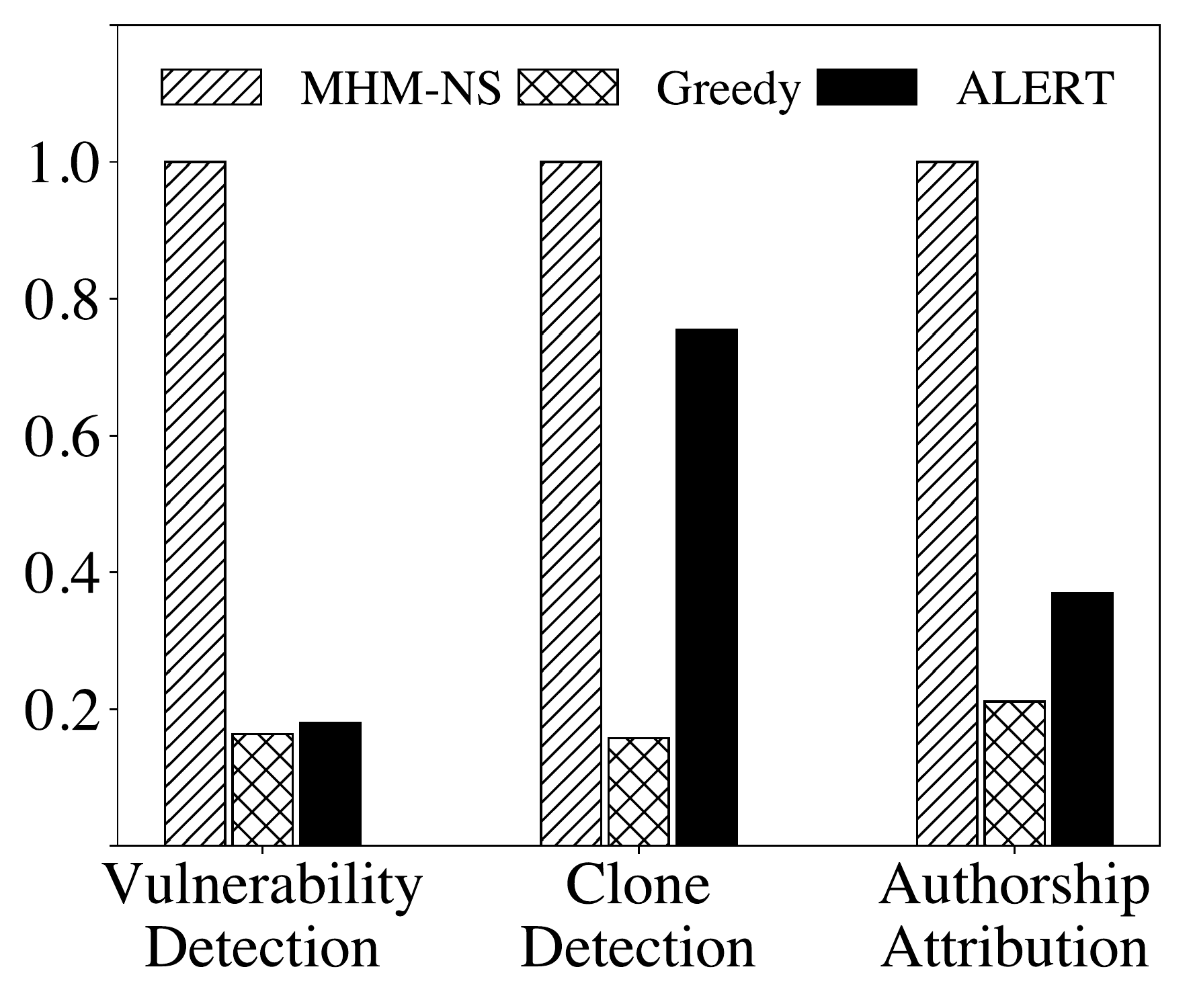}
    }
    \caption{Comparison results of Variable Change Rate (VCR) and Number Of Queries (NoQ) on attacking CodeBERT and GraphCodeBERT. The y-axis corresponds to the normalized values of VCR and NoQ. The x-axis represents downstream tasks.}
	\label{fig:3_metrics}
\end{figure*}

As the original MHM only works for code snippets written in C, we randomly sample some code snippets that can be successfully attacked by {\em ALERT}, MHM, and MHM-NS from the dataset of the vulnerability detection task (Section~\ref{subsec:dataset}). We have introduced a few more constraints when sampling the code snippets: (1) To save participants from reading long code snippets, we intentionally sample succinct and short code segments by limiting the code snippet length to 200 tokens; and (2) The attack methods may choose to replace different variables in the same code snippet; we only select the examples for which at least one variable is modified by all the three methods to make the comparison fair. There are 196 C code snippets that satisfy the aforementioned constraints. We compute a statistically representative sample size using a popular sample size calculator\footnote{\url{https://www.surveysystem.com/sscalc.htm}. Accessed: 2021-08-19} with a confidence level of 99\% and a confidence interval of 10. We sample 100 code snippets to conduct the user study, which is statistically representative.

%prepare four pairs copies (variants) of each selected code snippet to users: original input and adversarial examples generated by the three attacks. We then construct 4 pairs for each; each pair contain the original input and one of the 

For each selected code snippet, we can construct 3 pairs. Each pair contains the original code snippet and an adversarial example generated by either {\em ALERT}, MHM, or MHM-NS. We highlight the changed variables in each pair and present them to users. Users are asked to evaluate to what extent the substitutions are naturally fitting into the source code contexts. Given the statement: "The new variable name looks natural and preserves the original meaning", participants need to give scores on a 5-point Likert scale~\cite{joshi2015likert}, where 1 means strongly disagree and 5 means strongly agree, following the same settings used by Jin et al.~\cite{Jin_Jin_Zhou_Szolovits_2020}. Participants do not know which attack method produces which adversarial example in a pair.

The user study involves four non-author participants who have a Bachelor/Master degree in Computer Science with at least four years of experience in programming. Each participant evaluates the 100 pairs individually. We calculate the average ratings given to adversarial examples generated by each attack method per participant, and present the results in Figure~\ref{fig:user_study}. The x-axis distinguishes each participant, and the y-axis shows the average ratings. The results show that the usage of {\em ALERT}-generated substitutions can help generate much more natural adversarial examples. The four participants give average scores of close to 4 to adversarial examples generated by \toolname and slightly lower average scores to adversarial examples generated by MHM-NS; these indicate that participants perceive that the substitutions generated by these two methods are natural. Participants consistently give lower scores (1.86 on average) to examples generated by MHM, showing that they think the variable substitutions are unnatural.

% \textbf{Inter-rater reliability} The inter-rater reliability is an indicator of the reproducibility and reliability of the collected data. It gauges to what extent researchers can give consistent feedback. We used the Kendall coefficient \cite{kendall} of the agreement metric to determine the inter-rate reliability, which is a widely-used metric that seeks to quantify the degree of agreement among independently working researchers on the same data sample. The resultant Kendall's coefficient for all methods are around \hjd{0.7} which indicates a high and consistent distribution of scores among all s, proves that our user study has a highly reliable result. 

\begin{tcolorbox}
    \textbf{Answers to RQ1}: Participants consistently find that adversarial examples generated by ALERT (a naturalness-aware method) are natural while those generated by MHM (a naturalness-agnostic method) are unnatural.
    %give higher scores (3.95 on average) to examples obtained by selecting replacement from {\em ALERT}-generated substitutions than examples generated by MHM (2.18), which emphasizes that {\em ALERT}-generated substitutions help generated natural adversarial examples.
\end{tcolorbox}

% \begin{figure*}[t!]
%     \centering
%     \subfigure[Variable Change Rate (VCR)]{
%     \includegraphics[width=0.3\linewidth]{figures/pcr_cb.pdf}
%     \label{fig:}
%     }
%     \quad
%     \subfigure[Position Change Rate (PCR)]{
%         \includegraphics[width=0.3\linewidth]{figures/pcr_cb.pdf}
%     \label{fig:}
%     }
%     \quad
%     \subfigure[Number Of Queries (NoQ)]{
%         \includegraphics[width=0.3\linewidth]{figures/pcr_cb.pdf}
%     \label{fig:example7}
%     }
%     \caption{Results for other 3 VCR, PCR and NoQ on attacking GraphCodeBERT.}
% 	\label{fig:3_metrics_2}
% \end{figure*}

\subsection*{RQ2. How \textit{successful} and \textit{minimal} are the generated adversarial examples? How \textit{scalable} is the generation process?}
To answer this question, we evaluate the effectiveness of \toolname and MHM-NS on attacking CodeBERT and GraphCodeBERT considering three dimensions. Specifically, we use three metrics, each capturing one quality dimension,  to measure the performance of an adversarial example generation method. Each metric is defined based on a dataset $X$, where each element $x \in X$ is a code snippet that has at least one local variable and a victim model $M$ that can predict all examples in $X$ correctly. The three metrics are defined as follows.

\begin{itemize}[leftmargin=*]
    \item \textbf{Attack Success Rate} (ASR): The ASR of an adversarial example generation method is defined as $\frac{|\{x | x \in X  \wedge M(x') \neq M(x) \}|}{|X|}$, where $x'$ is a generated example. A higher ASR indicates that an attack method has better performance.
    \item \textbf{Variable Change Rate} (VCR): Assuming that an input code snippet $x_i$ has $m_i$ local variables, and an attacker renames $n_i$ variables in $x_i$, we define variable change rate (VCR) of the attack over $X$ as $\frac{\sum_{i} n_i}{\sum_{i} m_i}$. A lower VCR is preferable since it means that fewer edits are made to find successful adversarial examples. 
    % \item \textbf{Position Change Rate} (PCR): In a code snippet, one variable can appear in multiple positions. We need to change all occurences of a renamed variable to preserve operational semantics. Assuming that an original example $x_i$ has $p_i$ occurrences in the code snippet, and an attacker renames $n_i$ variable occurrences, we define position change rate (PCR) of the attacker over $X$ as $\frac{\sum_{i} n_i}{\sum_{i} p_i}$. PCR is also a supplementary metric, which is flavoured for a lower value.
    \item \textbf{Number of Queries} (NoQ): In adversarial attacks, especially the black-box ones, the number of queries to the victim model needs to be kept as low as possible. In practice, victims models are usually remotely deployed, and it is expensive (and maybe also suspicious) to query models too many times. We count the number of queries (NoQ) to the victim models when each attack generates adversarial examples on dataset $X$. Attacks that have lower NoQ are more scalable as well.
\end{itemize}

Table \ref{tab:ASR} displays the comparison results between MHM-NS and \toolname on the six victim models (2 models $\times$ 3 tasks as described in Section~\ref{subsec:dataset}). We also report the results of solely using Greedy-Attack to emphasize the improvements brought by GA-Attack. Results show that Greedy-Attack has 49.42\%, 23.20\% and 30.28\% attack success rate on CodeBERT across three downstream tasks, which corresponds to an improvement of 13.76\%, 15.71\% and 11.01\% over MHM-NS, respectively. By employing GA-Attack in {\em ALERT}, we can boost the performance even further: MHM-NS results are improved by 17.96\%, 7.74\% and 16.51\% in terms of ASR. On GraphCodeBERT, Greedy-Attack can outperform MHM-NS by 16.81\%, 3.33\% and 14.68\% for the three tasks; the numbers are boosted to 21.78\%, 4.54\% and 29.36\% when GA-Attack is employed.

% Looking at the results and approaches more closely, we find that attacking GraphCodeBERT on Clone Detection task get worst performance since MHM and our methods all use variable renaming to generate adversarial examples and the fine-tuned model on this task is not sensitive to variable changes. However, Greedy-Attack and \toolname still get decent relative improvements over MHM-NS on this hardest task as the application of importance score can benefit the adversarial examples generation and make better use of model output than MHM does at the same setting. In addition to Clone Detection task, all methods on attacking GraphCodeBERT get a better success rate than that on attacking CodeBERT, which indicates that GraphCodeBERT is more vulnerable than CodeBERT across these downstream tasks.

Moreover, \toolname makes fewer edits to the original examples and is more scalable than the baseline. Figure~\ref{fig:3_metrics} compares results in terms of VCR and NoQ. The x-axis corresponds to each downstream task, and the y-axis represents normalized values of the two evaluation metrics. On all victim models, \toolname modifies fewer variables to generate adversarial examples. It indicates that \toolname can make minimal changes to input code snippets and produce more natural and imperceptible adversarial examples. Besides, \toolname queries victim models less than MHM-NS does. The NoQ of solely using Greedy-Attack is 82.57\% less than MHM-NS. When GA-Attack is employed, the NoQ increases but is still 49.62\% less than MHM-NS, which shows that \toolname is more practical since victim models are usually remotely deployed and may be costly to query and may prevent frequent queries. Querying victim models is the most time-consuming part of experiments, so fewer NoQ also shows that \toolname has lower runtime.

\begin{tcolorbox}
    \textbf{Answers to RQ2}: In terms of the attack success rate, \toolname can outperform the MHM by 17.96\%, 7.74\% and 16.51\% on CodeBERT, as well as 21.78\%, 4.54\% and 29.36\% on GraphCodeBERT across three downstream tasks. In addition to achieving a superior attack success rate, our method also makes fewer changes and is more scalable.
\end{tcolorbox}

\begin{table*}[!t]
    \caption{Robustness analysis on adversarially fine-tuned victim models. The numbers are the prediction accuraies of adversarially fine-tuned models (CodeBERT-Adv and GraphCodeBERT-Adv) on the adversarial examples generated in RQ2.}
    \begin{tabular}{lllllll}
        \hline
        \multirow{2}{*}{Tasks}  & \multicolumn{3}{l}{CodeBERT-Adv}  & \multicolumn{3}{l}{GraphCodeBERT-Adv} \\ \cline{2-7} 
                                & MHM-NS     & Greedy  & \toolname & MHM-NS       & Greedy   & \toolname  \\ \hline
        Vulnerability Detection & 80.46\% & 87.93\% & 88.11\%   & 80.81\%   & 88.84\%  & 89.04\%    \\
        Clone Detection         & 59.33\% & 91.38\% & 87.31\%   & 48.28\%   & 91.23\%  & 91.70\%    \\
        Authorship Attribution  & 63.89\% & 83.97\% & 87.36\%   & 98.72\%   & 96.40\%  & 96.21\%    \\ \hline
        Overall  & 67.89\% & 87.76\% & 87.59\%   & 75.93\%   & 92.16\%  & 92.32\%
        \\ \hline
        \end{tabular}
    \label{tab:adv_training}
    \end{table*}

\subsection*{RQ3. Can we use adversarial examples to harden the victim models?}
In this research question, we explore the effectiveness of using adversarial fine-tuning \cite{hosseini2017limitation} as a defense against attacks. We leverage \toolname to generate adversarial examples for each victim model on their corresponding training sets. If a victim model predicts wrongly on an original input or no local variable name can be extracted from it, we skip this example. For other inputs in the training sets, we select at most one adversarial example for each of them. If \toolname attacks successfully, we choose the firstly-found adversarial example. If \toolname fails to attack, we select the example that can minimize the victim model's confidence on the ground truth label. These generated adversarial examples are then augmented into the original training set and form the \textit{adversarial training set}. We then fine-tune the victim model on the adversarial training set.

After adversarial fine-tuning, we obtain two models: CodeBERT-Adv and GraphCodeBERT-Adv and evaluate them on the adversarial examples generated in RQ2. Table~\ref{tab:adv_training} shows the new models' prediction accuracy on previously generated adversarial examples. It is noted that the original victim models (that are not hardened by adversarial retraining) predict all these examples wrongly (i.e., they have an accuracy of 0\%). From Table~\ref{tab:adv_training}, we can observe that all the adversarially fine-tuned models perform much better than the original ones. The average improvement on examples generated by solely using Greedy-Attack and employing GA-Attack is close. CodeBERT-Adv improves accuracy against Greedy-Attack and \toolname by 87.76\% and 87.59\%, respectively. GraphCodeBERT-Adv improves accuracy against Greedy-Attack and \toolname by 92.16\% and 93.31\%. Accuracy improvement on adversarial examples generated by MHM-NS is relatively more minor (67.89\% and 75.93\% on CodeBERT and GraphCodeBERT). 

% , models robustness are improved against \emph{ALERT}, Greedy-Attack in \toolname and MHM-NS by 87.59\%, 87.76\% and 67.89\% on average. On GraphCodeBERT, adversarial fine-tuning can enhance models robustness by 92.31\%, 92.16\% and 75.93\% on average against {\em ALERT}, Greedy-Attack in \toolname and MHM-NS. 
%However, the adversarial fine-tuning models perform not well on the adversarial examples generated by MHM-NS since the VCR of MHM-NS as shown in RQ2 is very high which means that MHM-NS makes huge changes into the original input code snippets as adversarial examples. These adversarial examples can be far beyond the data distribution of original training and adversarial training set.

\begin{tcolorbox}
    \textbf{Answers to RQ3}: The adversarial examples generated by \toolname are valuable in improving the robustness of victim models. Adversarially fine-tuning victim models with {\em ALERT}-generated adversarial examples can improve the accuracy of CodeBERT and GraphCodeBERT by 87.59\% and 92.32\%, respectively.
    \end{tcolorbox}

%% file: sections/discussion.tex
% \section{Discussion}

\section{Threats to Validity}
\label{sec:ttv}
\vspace{0.2cm} \noindent{\bf Internal validity:} 
The results obtained in our experiment can vary under different hyper-parameters settings, e.g., input length, numbers of training epochs, etc. To mitigate the threats, we set the input length to CodeBERT and GraphCodeBERT as 512 (the maximal value) to ensure that they see the same numbers of tokens for the same code snippet. For the remaining hyper-parameters, we keep them the same as described in~\cite{CodeBERT, GraphCodeBERT}. We compare the performance of models obtained in this paper with results reported in the literature~\cite{CodeBERT, GraphCodeBERT, CodeXGLUE} to show that our models are properly trained.

\vspace{0.2cm} \noindent{\bf External validity:} 
In our experiments, we investigate two popular pre-trained models of code on three downstream tasks. However, our results may not generalize to other pre-trained models and downstream tasks. We use a generic parser to extract variable names from code snippets written in C, Python or Java, but it cannot work in other programming languages like Ruby.

% \vspace{0.2cm} \noindent{\bf Construct validity:} 
% In our experiments, we investigate two popular pre-trained models of code on three downstream tasks. However, our results may not generalize to other pre-trained models and downstream tasks. We use a generic parser to extract variable names from code snippets written in C, Python or Java, but it cannot work in other programming languages like Ruby.

%% file: sections/related_work.tex
\section{Related Work}
\label{sec:related_work}
This section describes the works that are related to this paper, including the pre-trained models of code and adversarial attacks on models of code.
\subsection{Pre-trained Models of Code}

Code representation models like code2vec~\cite{code2vec} and code2seq~\cite{code2seq} that use syntactic and structural information have shown good performance on a range of downstream tasks.
However, some pre-trained models for Natural Languages (NL) like BERT~\cite{bert} and GPT-3~\cite{GPT-3} have recently demonstrated excellent transferability to Programming Languages (PL) and stronger capabilities of capturing semantics information than code2vec or code2seq. Inspired by the success of these language models, pre-trained models of code have recently become more and more popular in the field of code intelligence and benefited a broad range of tasks~\cite{CodeBERT, GraphCodeBERT, CuBERT, CBERT, svyatkovskiy2020intellicode, wang2021codet5}.
These current pre-trained models of code can be divided into two types: embedding models and generative models. 

The two models (CodeBERT~\cite{CodeBERT} and GraphCodeBERT~\cite{GraphCodeBERT}) investigated in our experiments are representatives of embedding models. 
We have described CodeBERT and GraphCodeBERT in Section~\ref{subsec:pre_train_models}. Here, we briefly describe other embedding models. Kanade et al.~\cite{CuBERT} used the same model architecture and training objectives as BERT but trained it on Python source code to produce CuBERT. Buratti et al.~\cite{CBERT} introduced C-BERT, a transformer-based language model trained on 100 popular C language repositories on Github. 
Both CuBERT and C-BERT were trained on a single programming language, which limits their usage scenarios. CuBERT and C-BERT outperform generic baselines like LSTM models but do not show superior performance than CodeBERT and GraphCodeBERT, so we investigate the latter two in this work. 

The other branch of pre-trained models is generative models, which are designed for generative tasks like code completion. Svyatkovskiy et al.~\cite{svyatkovskiy2020intellicode} introduce GPT-C, a variant of GPT-2~\cite{GPT-2} trained on a large corpus containing multiple programming languages and achieved impressive performance in code generation tasks. Lu et al.~\cite{CodeXGLUE} provides CodeGPT, which has the same model architecture and training objective of GPT-2. CodeGPT was trained on Python and Java corpora from the CodeSearchNet dataset~\cite{CodeSearchNet}. Despite their success on generation tasks, these models are unable to get complete contextual information as they are unidirectional decoder-only models which only rely on previous tokens and ignore the following ones~\cite{wang2021codet5}, so we discard generative models in the investigation list of this work.

% Most pre-trained programming languages models that inherit the BERT architecture would include MLM, NSP or RTD tasks as the training objectives. However, these pre-trained tasks do not treat code differently from natural languages. Roziere et al. \cite{roziere2021dobf} has introduced a deobfuscation pre-training objective that is explicitly designed for programming languages. Ahmad et al. \cite{ahmad2021unified} PLBART is another pre-trained model train on the checkpoints of BART \cite{lewis2019bart} with an extensive collection of Java and Python functions collected From Github and StackOverflow.

% \noindent \textbf{Robustness of DNNs}:

\subsection{Adversarial Attack on Models of Code}
Yefet et al. \cite{while_box_attack_code} proposed DAMP, a white-box attack technique that adversarially changes variables in code using gradient information of the victim model. Although their method shows effectiveness in attacking three models {\em code2vec} \cite{code2vec}, {\em GGNN} \cite{allamanis2018learning}, and {\em GNN-FiLM} \cite{brockschmidt2018generative}, it requires victim models to process code snippets using one-hot encoding, which is not applicable to CodeBERT~\cite{CodeBERT} and GraphCodeBERT~\cite{GraphCodeBERT} investigated in our paper as they use BPE~\cite{Gage1994ANA, RoBERTa} to process tokens. Srikant et al.~\cite{iclr2021} apply PGD~\cite{PGD} to generate adversarial examples of code. 
Besides, these white-box approaches generate substitutes by changing a one-hot encoding to another and mapping it back to a token, which cannot guarantee to satisfy naturalness requirements. Such a white-box attack is less practical since victim models are usually deployed remotely, making parameter information hard to be accessed. 

There are several black-box methods for evaluating the robustness of models of code. One that has been shown to be much more effective than the others is MHM~\cite{MHM}, which we use as our baseline. We have presented the details of MHM in Section~\ref{subsec:mhm}. Here we present the other related studies. Wang et al.~\cite{wang2019coset} provide a benchmark consisting of refactored programs and evaluate the performance of neural embedding programs on it. Rabin et al.~\cite{rabin2021generalizability} also uses variable renaming to evaluate the generalizability of neural program analyzers, and show that GGNN \cite{fernandes2018structured} changes its prediction on 33.39\% of transformed code. Pour et al. \cite{9438605} proposed a testing framework for DNN of source code embedding, which can decrease the performance of code2vec \cite{code2vec} on method name prediction task by 2.05\%. 
Applis et al.~\cite{ase_nier} use metamorphic program transformations to assess the robustness of ML-based program analysis tools in a black-box manner. 

Adversarial attack on models of code can be conducted beyond generating adversarial examples for a well-trained model. Schuster et al.~\cite{263874} show that code completion models are vulnerable to poisoning attacks that add some carefully-designed files to the training data of a model. Nguyen et al.~\cite{ase_smell} show that the state-of-the-art API recommender systems can be attacked by injecting malicious data into their training corpus. 

%% file: sections/conclusion.tex
\section{Conclusion and Future Work}
\label{sec:conclusion}
In this paper, we highlight the naturalness requirement in generating adversarial examples for models of code. We propose \toolname (N\textbf{a}tura\textbf{l}n\textbf{e}ss Awa\textbf{r}e A\textbf{t}tack), a black-box attack that adversarially transforms inputs (code snippets) to force pre-trained models to produce wrong outputs. \toolname can generate naturalness-aware substitutes. A user study confirms that these substitutes can help generate adversarial examples that look natural to human judges. In contrast, users consistently think examples generated by a prior method that employs random selection to be unnatural. Apart from being aware of naturalness, \toolname is also effective in finding adversarial examples. We apply \toolname to victim models fine-tuned on state-of-the-art pre-trained models (CodeBERT and GraphCodeBERT). The results show that on attacking CodeBERT, \toolname can achieve average success rates of 53.62\%, 27.79\%, and 35.78\% across three downstream tasks: vulnerability prediction, clone detection and code authorship attribution. It outperforms the baseline by 17.96\%, 7.74\% and 16.51\%. On GraphCodeBERT, our approach can achieve average success rates of 76.95\%, 7.96\% and 61.47\% on the three tasks, respectively, outperforming the baseline by 21.78\%, 4.54\%, and 29.36\%. We also explore the value of adversarial examples to harden CodeBERT and GraphCodeBERT through an adversarial fine-tuning procedure and demonstrated the robustness of CodeBERT and GraphCodeBERT against \toolname increased by 87.59\% and 92.32\%, respectively. We open-source \toolname at \url{https://github.com/soarsmu/attack-pretrain-models-of-code}.

In the future, we plan to consider more victim models and more downstream tasks. We also plan to boost the effectiveness of \toolname and improve the robustness of victim models further. %in terms of the various quality dimensions.